\RequirePackage{fix-cm}
\documentclass[twocolumn]{svjour3}          % twocolumn
\smartqed  % flush right qed marks, e.g. at end of proof

\usepackage{cite}
\usepackage[pdftex]{graphicx}
\usepackage{ragged2e}
\usepackage[tight,footnotesize]{subfigure}
\usepackage{rotating}
\usepackage{graphicx}
\usepackage{amsmath,amssymb} % define this before the line numbering.
\usepackage{array}
\usepackage{multirow}
\usepackage{colortbl}
\usepackage{amsfonts}
\usepackage{pifont}
\usepackage{xspace}
\usepackage{etoolbox}
\usepackage{overpic}
\usepackage{color}
\usepackage{microtype}
\usepackage{xcolor}

\newcommand{\orcid}[1]{\href{https://orcid.org/#1}{\includegraphics[width=10pt]{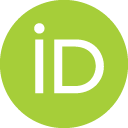}}}

\usepackage{hyperref}
\hypersetup{breaklinks=true,citecolor=blue, colorlinks}

\graphicspath{{./}}
\DeclareGraphicsExtensions{.pdf,.jpg,.png}

\usepackage{silence}
\journalname{Research Article}

\begin{document}

\renewcommand{\figurename}{Figure}
\renewcommand{\tablename}{Table}

\title{TVC: Tokenized Video Compression with Ultra-Low Bit Rate}

% \titlerunning{Short form of title}        % For running head
\titlerunning{TVC}

\author{
  Lebin Zhou \orcid{0009-0008-1167-3098}       \and
  Cihan Ruan \orcid{0009-0006-3094-0505}       \and 
  Nam Ling \orcid{0000-0002-5741-7937}         \and
  Zhenghao Chen \orcid{0000-0003-0155-4462}    \and
  Wei Wang \orcid{0000-0002-3765-5241}         \and
  Wei Jiang \orcid{0000-0001-6672-9783}     
}

% \authorrunning{F. Author \etal} % if too long for running head
\authorrunning{Zhou et al.}

% \institute{
% First Author and Second Author are with institution, (department), city, (state), country. 
% (Email: xxx@xxx.xx.xx, xxx@xxx.xx.xx). \\
% Third Author is with institution, (department), city, (state), country. 
% (Email: xxx@xxx.xx.xx). \\
% Corresponding author: XXXXXXX.
% }

\institute{
Lebin Zhou, Cihan Ruan and Nam Ling are with Santa Clara University, Santa Clara, CA, USA. 
(Email: lzhou@scu.edu, cruan@scu.edu, nling@scu.edu). \\
Wei Wang and Wei Jiang are with Futurewei Technologies, Inc., San Jose, CA, USA. 
(Email: rickweiwang@futurewei.com, wjiang@futurewei.com). \\
Zhenghao Chen is with University of Newcastle, Callaghan, NSW, Australia.
(Email: zhenghao.chen@newcastle.edu.au). \\
Corresponding author: Lebin Zhou(lzhou@scu.edu).
}

\date{Received: date / Accepted: date}

\maketitle

\begin{abstract}
Tokenized visual representations have shown promise in image compression, yet their extension to video remains underexplored due to the challenges posed by complex temporal dynamics and stringent bit rate constraints. In this paper, we present tokenized video compression (TVC), a token-based dual-stream framework designed to operate effectively at ultra-low bit rates. TVC leverages the Cosmos video tokenizer to extract both discrete and continuous token streams. The discrete tokens are partially masked using a strategic masking scheme and then compressed losslessly with a discrete checkerboard context model to reduce transmission overhead. The masked tokens are reconstructed by a decoder-only Transformer with spatiotemporal token prediction. In parallel, the continuous tokens are quantized and compressed using a continuous checkerboard context model, providing complementary continuous information at ultra-low bit rates. At the decoder side, the two streams are fused with a ControlNet-based multi-scale integration module, ensuring high perceptual quality alongside stable fidelity in reconstruction. Overall, this work illustrates the practicality of tokenized video compression and points to new directions for semantics-aware, token-native approaches.
% Please provide 4 to 6 keywords which can be used for indexing purposes.
% \keywords{First keyword \and Second keyword \and More}


\end{abstract}

\section{Introduction}
\label{sec:intro}

Generative visual priors, also known as visual tokens, have demonstrated remarkable efficacy in various restoration tasks, such as super-resolution \cite{Adacode,FemarSR,CosmoSR}, quality enhancement \cite{VQDeblur}, and image compression \cite{Hybridflow,VQGANImage}. These visual priors are trained on a substantial corpus of images and videos to model the distributions of the entire visual space. The key is the learned tokenized visual representation (TVR). A tokenizer transforms the input visual signals into the TVR, which a pixel decoder then uses to reconstruct the output visual signals. The tokenizer, TVR, and pixel decoder are optimized end-to-end to strike a balance between representation efficiency and reconstruction quality.

\begin{figure}[t]
  \includegraphics[width=1.0\linewidth]{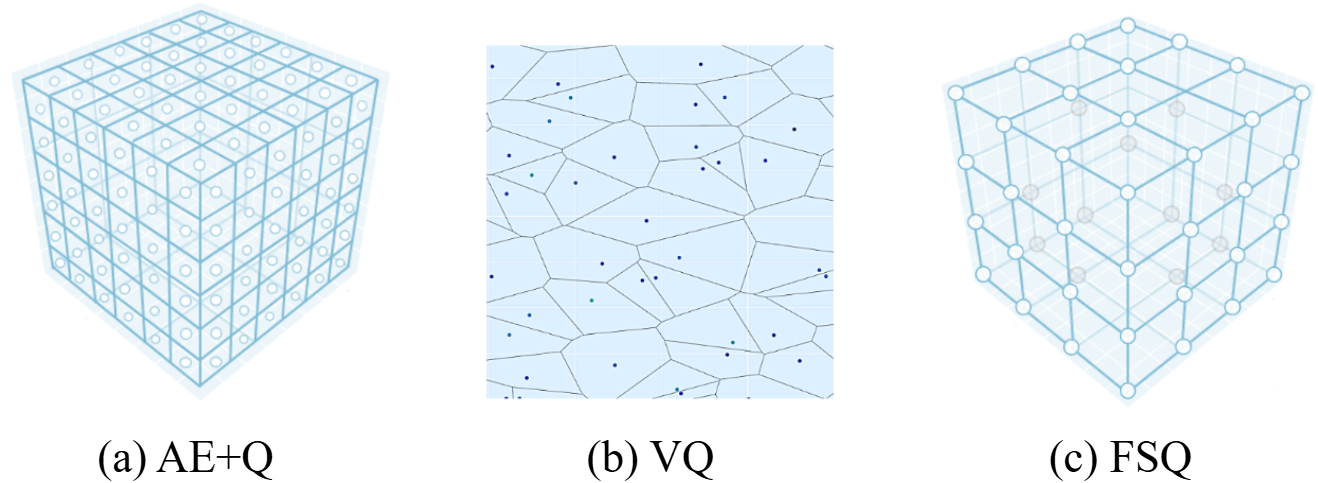}
  \caption{Visualization of quantization strategies of different tokenized visual representations (TVRs). Each cell represents a token, with the dot indicating its centroid (latent feature). The AE+Q represents applying scalar quantization (Q) to the latent features from the autoencoder (AE). This results in a uniform and equal partition of the latent space. In contrast, the non-uniform partition of vector quantization (VQ) and finite scalar quantization (FSQ) emphasizes frequent visual patterns and enables adaptive representation, leading to higher perceptual quality under extreme compression.}
  \label{fig:TVRtoy}
  \vspace{-1.0em}
\end{figure}

There are two types of TVR: continuous TVR (C-TVR) and discrete TVR (D-TVR). Continuous tokenizers , e.g., autoencoder (AE), variational autoencoder (VAE) \cite{VAE,CosmoSR}, transform inputs into a learned latent space as continuous latent features, which are then quantized for transmission. The quantization process partitions the latent space of C-TVR into equal cells, as illustrated in Fig.~\ref{fig:TVRtoy}(a). The discrete tokenizers , e.g., vector quantized generative adversarial network (VQGAN)  \cite{VQGANImage,Magvit,MAGE} and finite scalar quantization (FSQ) \cite{FSQ,CosmoSR}, transform inputs into discretized sequences of latent codes by learning a latent visual space that is partitioned into cells with unequal volumes, as illustrated in Fig.~\ref{fig:TVRtoy}(b) and (c). 

C-TVR has been largely studied for learned image compression (LIC) \cite{chen2020,MLIC}, since the equal partition of the visual space preserves the local sensitivity of the visual features. Similar and different visual appearances are assigned to similar and different cells, respectively, in latent space, and the granularity of partition, i.e., the scale of quantization, determines the mapping sensitivity and, consequently, the fidelity of the reconstruction to the input. This aligns with the conventional compression target of balancing rate-distortion (RD). However, at ultra-low bit rates, the performance suffers from severe information loss caused by heavy quantization, as is the case for conventional compression methods.

D-TVR has gained increasing attention in recent years for low bit rate LIC \cite{VQGANImage,Hybridflow} due to its ability to generate reconstruction with high perceptual quality against input degradations. The learned unequal partitions focus on modeling the general salient visual cues described by dense areas in the visual space, while allocating sparse cells over less representative areas. Such a ``smart" token allocation is especially useful for ultra-low bit rate scenarios, where outputs can be reconstructed with high perceptual quality by using only a small number of salient tokens.

Neither D-TVR nor C-TVR has been applied to compress videos, although there is a much stronger need for ultra-low bit rate video compression compared to images, since efficient video transmission has become a performance bottleneck for a vast amount of applications in entertainment, gaming, AR/VR , etc. The survey \cite{ma2023overview} shows that previous video compression methods, including learned video compression (LVC) based on neural networks \cite{DVC,FVC}, follow a pipeline based on motion prediction and residual coding, which was designed decades ago, particularly for traditional video coding \cite{VVC,HEVC,MPEG}. We strongly believe that the overall performance of LVC, in terms of both computation and bit rate, can be largely improved by redesigning a holistic framework based on TVRs. However, there is a fundamental dilemma about using TVR for LVC. On the one hand, to reliably represent the complex spatial-temporal visual content in general videos, D-TVR usually requires a very large number of spatial-temporal tokens (tens or even hundreds of millions), and C-TVR usually requires a latent space with very high dimensionality. The number of bits to represent such token indices or high-dimensional latent features can be prohibitively high, defeating the purpose of video compression.

\begin{figure*}[t]
  \includegraphics[width=1.0\linewidth]{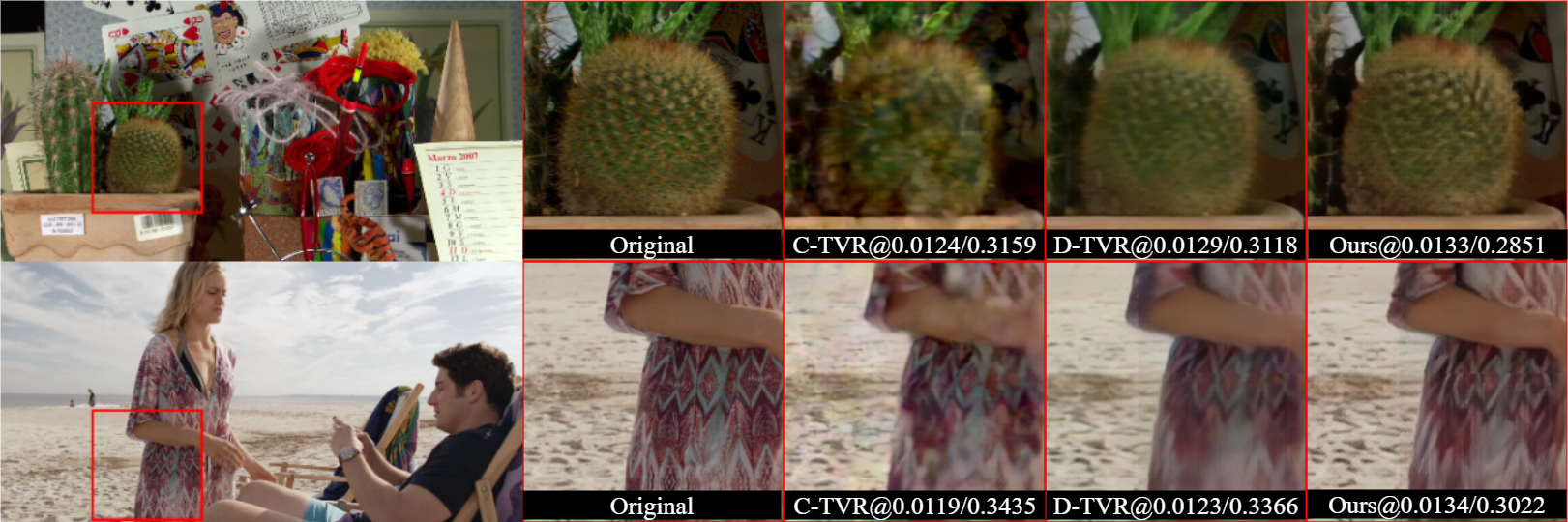}
  \caption{Visual comparison between single-stream and dual-stream reconstructions at extremely low bit rates. The first example is from HEVC-B\cite{HEVC}, and the second example is from MCL-JVC\cite{MCL-JCV}. The corresponding bpp/LPIPS values are reported. C-TVR produces overly smooth and blurry outputs as a result of heavy quantization, while D-TVR often hallucinates inauthentic details derived from learned salient codewords. Our dual-stream design combines their complementary strengths and faithfully recovers both rich textures and structural details.}
  \label{fig:single_stream}
  \vspace{-0.5em}
\end{figure*}

\begin{figure*}[t]
  \includegraphics[width=1.0\linewidth]{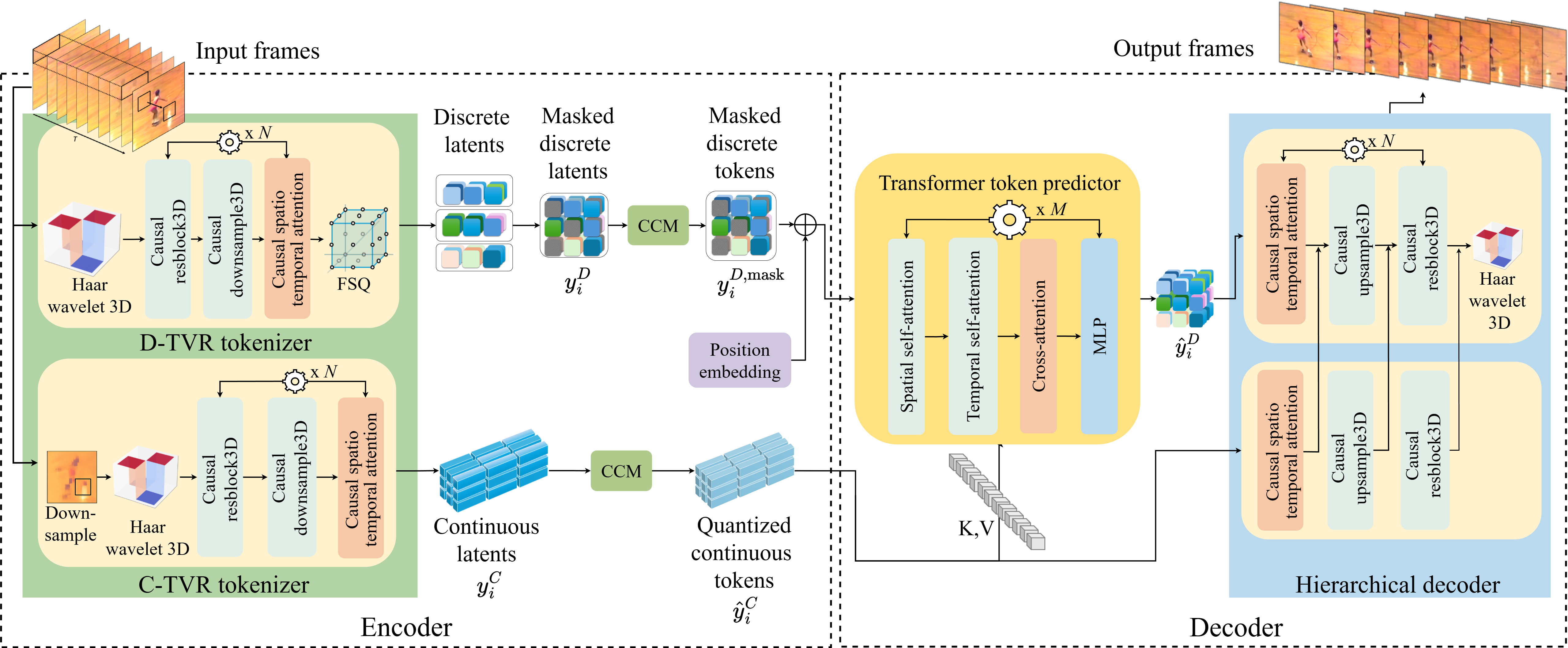}
  \caption{
    Overview of the proposed tokenized video compression (TVC) framework.
    The tokenizer encoder extracts discrete and continuous token streams from each input GoP. Discrete tokens are masked and compressed using a checkerboard context model (CCM), and continuous tokens are quantized and also compressed with CCM. At the decoder side, masked discrete tokens are reconstructed via a transformer with cross-attention to the continuous stream. Both streams are fused via a ControlNet-style\cite{zhang2023controlnet} hierarchical pixel decoder for final frame reconstruction. MLP: multilayer perceptron; K, V: key and value inputs for cross-attention, respectively; $y^{D}_{i}$: discrete latent token; $y^{D,\text{mask}}_{i}$: masked discrete token; $y^{C}_{i}$: continuous latent token; $\hat{y}^{C}_{i}$: quantized continuous token; $\hat{y}^{D}_{i}$: predicted discrete token; $N$: the number of 3D causal spatio temporal blocks; $M$: the number of transformer layers in the token predictor.
  }
  \label{fig:network_structure}
  \vspace{-1.1em}
\end{figure*}

In this paper, we introduce a novel dual-stream tokenized video compression (TVC) framework, which is systematically designed for TVR-based LVC, as illustrated in Fig.~\ref{fig:network_structure}. Our approach addresses the aforementioned challenges from several aspects and integrates both D-TVR and C-TVR to leverage their complementary strengths. The key contributions of our work are summarized as follows.

\begin{itemize}
    \item[1)] We aim for both high fidelity and high perceptual quality in reconstructed videos at ultra-low bit rates. Although D-TVR excels at high-quality visual generation, it can result in loss of fine-fidelity details. Visual differences in the sparse regions of the latent space can be overlooked, while inauthentic details can be hallucinated in denser areas (as illustrated by examples in Fig.~\ref{fig:single_stream}. By integrating the fidelity-preserving C-TVR stream, we achieve a balanced outcome that ensures both high fidelity and enhanced perceptual quality.\vspace{.2em}

    \item[2)] We take advantage of the high redundancy in videos to largely reduce the number of transferred tokens by employing the token prediction mechanism. As shown by masked generative encoder (MAGE) \cite{MAGE}, the token space creates an efficient probability space for prediction, where masked token prediction can effectively recover image content using spatial contextual information. Compared to images, token-based modeling is even more effective for videos, since semantic patterns tend to persist between frames \cite{dosovitskiy2020image}. Tokens capture content at a conceptual level where semantic consistency naturally emerges, i.e., similar visual concepts in adjacent frames often map to identical or closely related tokens, creating temporal predictable patterns. Our TVC exploits this semantic redundancy through spatiotemporal token prediction to achieve efficient compression.    \vspace{.2em}

    \item[3)] We build upon Nvidia’s Cosmos tokenizer \cite{CosmoSR}, which extracts compact discrete and continuous token streams through wavelet transforms, spatiotemporal 3D convolution, and spatiotemporal causal self-attention. To effectively further compress these dual streams, we introduce two checkerboard context models (CCM). For discrete tokens, we apply 3D masking to drop off masked tokens and losslessly compress the visible tokens using a discrete CCM. For continuous tokens, we perform quantization with a continuous CCM for effective lossy compression. This design helps reduce both spatial and temporal redundancy, enabling operation at ultra-low bit rates. \vspace{.2em}

    \item[4)] Inspired by the residual conditioning mechanism of ControlNet \cite{zhang2023controlnet}, we design a hierarchical pixel decoder that fuses D-TVR and C-TVR streams through multi-scale residual injection. Specifically, the C-TVR latents are decoded via a cascade of Cosmos pixel decoder layers, whose intermediate features are injected into the corresponding layers of the D-TVR pixel decoder in a residual manner. This dual-stream fusion architecture enhances both pixel-level fidelity and semantic consistency across video frames.
    
\end{itemize}

Compared to conventional LVC methods such as Refs.~\cite{DVC,FVC}, our TVC approach significantly improves ultra-low bit rate performance with considerably reduced computational overhead. Likewise, compared to the straightforward extension of dual-stream LIC methods such as Ref.~\cite{lu2025hdcompression}, our TVC consistently outperforms by leveraging effective video tokenizers and enhancing the reduction of spatio-temporal redundancy through advanced spatio-temporal token prediction.

Operating in the token space, we formulate video compression as a token selection and prediction problem. Our key insight is that, with a structured and temporally causal token space like Cosmos, ultra-low bit rate video compression becomes not only feasible but highly effective. Unlike prior methods that focus on compressing code indices or residuals, we directly compress the tokens themselves, yielding a highly compact and semantically consistent representation across frames. Our approach addresses the challenge about the practicality of tokenized video compression and suggests new opportunities for learned video compression.

It is worth mentioning that our TVC framework is designed with flexibility in mind, allowing seamless integration with different video tokenizers beyond Cosmos. This adaptability stems from our generic token prediction and fusion mechanisms, which are agnostic to any specific tokenizer architecture.

We conduct extensive experiments and ablation studies on multiple benchmark datasets. As the first tokenized video compression system operating at ultra-low bit rates, TVC achieves an LPIPS of around 0.30 without relying on conventional optical flow or motion estimation and motion compensation (MEMC). At such extreme bit rate levels, pixel-level distortion metrics like PSNR and perceptual quality metrics like LPIPS often diverge, highlighting the trade-off between structural fidelity and visual realism. In this context, LPIPS serves as a more reliable indicator of perceptual quality, aligning more closely with human visual preferences. Our results exhibit strong perceptual consistency across frames, demonstrating the potential of token-based modeling for semantic-level video reconstruction under extreme compression constraints.

\section{Related Works}
\label{sec:relatedWorks}

\subsection{Tokenized Visual Representation}
\label{sec:relatedTVR}

The core of image/video generation is the learned TVR by modeling the probability distributions of the visual space. For D-TVR, the discrete tokenizer, e.g., VQGAN \cite{VQGANImage} or FSQ \cite{FSQ}, transforms input visual signals into discretized sequences of latent codes. For C-TVR, the continuous tokenizer, e.g., VAE \cite{VAE,CosmoSR}, transforms inputs into a learned latent space as continuous latent features. Then a pixel decoder transforms TVR back to the pixel space to reconstruct the visual signal. The tokenizer, TVR, and pixel decoder are optimized end-to-end to balance the efficiency of representation and the reconstruction quality. 

The spatiotemporal visual content in videos is inherently complex, and most existing tokenizers focus solely on spatial image content \cite{VQGANImage,FSQ,VAE}. Even for static images, developing an efficient class-agnostic tokenizer capable of handling diverse image categories remains challenging. Improvements have been made through class-guided tokenizers. For instance, AdaCode \cite{Adacode} introduces a set of basis codebooks to capture semantic-specific visual details and employs a weight map to effectively combine these basis codebooks.

However, directly applying image-based TVR to tokenize individual video frames introduces temporal inconsistencies, resulting in jittering and flickering artifacts. To address this, video-based TVRs \cite{omnitoken,CosmoSR} have been developed by training on extensive datasets comprising tens of thousands or even millions of hours of video content. Given the heightened complexity of spatiotemporal visual content, video-based TVRs typically require a significantly larger number of spatiotemporal tokens or a highly dimensional latent space for accurate modeling compared to their image-based counterparts.

\subsection{Tokenization for Image Compression}
\label{sec:image_compression}

Since the pioneer work of Ballé et al. \cite{balle2018}, C-TVR based on VAE has emerged as a prominent approach in LIC. In this method, a continuous tokenizer transforms an image into a latent feature, which is then processed through traditional quantization and entropy coding to produce a compact bitstream with continuous values. The decoder reconstructs the image by applying conventional entropy decoding and dequantization to recover the degraded latent feature. While significant advancements have been made in improving the entropy model to mitigate information loss during quantization, achieving high-quality reconstruction at ultra-low bit rates remains challenging, because aggressive quantization severely degrades the recovered latent feature, often resulting in artifacts such as blurring and blocky effects.

The idea of D-TVR naturally aligns with compression tasks, where an image is mapped to a sequence of discretized latent codes. Each code corresponds to the index of a partitioned cell, i.e., a visual token, in the learned latent space. Using the index, the decoder retrieves the latent feature of the corresponding cell for reconstruction. Compared to the original embedded feature, the retrieved latent feature of the mapped token is a quantized approximation. The granularity of this partition,  i.e. the number of tokens, directly determines the tradeoff between bit rate and reconstruction quality. A finer partition with more tokens allows for higher reconstruction quality at the cost of increased bit rate. 

D-TVR offers two notable advantages for compression. First, the use of integer indices is highly efficient for data transmission, ensuring robustness across different platforms and minimizing computation mismatches between sender and receiver devices. Second, D-TVR can achieve high perceptual quality in reconstructed content by leveraging high-quality tokens. Even when the input quality is degraded, retrieving a set of visual tokens similar to those representing a high-quality input can significantly enhance the reconstruction outcome. This capability is particularly advantageous in ultra-low bit rate scenarios, where C-TVR or traditional methods often struggle to deliver high-quality results. 

Therefore, D-TVR has been used recently for ultra-low bit rate LIC. 
Since learning a universally rich visual codebook capable of capturing diverse image content remains very challenging, the direct application of VQGAN to LIC \cite{LICVQGAN} has demonstrated improvements in perceptual quality at extremely low bit rates. To enhance performance across a broader range of bit rates, Ref.~\cite{maskSVR} introduced multiple visual codebooks that capture class-specific visual details \cite{Adacode}, combined with a weight masking mechanism to mitigate the increased bit rate associated with transmitting the additional weights for fusing multiple codebooks. Despite the advancements, D-TVR alone still faces the inherent fidelity loss issue, particularly at ultra-low bit rates. In such cases, a smaller codebook excessively compresses the visual space, causing different images to be treated as variations of one another and mapped to the same set of codewords. This results in reconstructions that, while potentially visually appealing, may lack pixel-level — and even semantic-level — accuracy, failing to faithfully reproduce the original input.

To overcome the inherent limitations of both the C-TVR and D-TVR methods, researchers have proposed dual-stream frameworks that leverage their complementary strengths. HybridFlow \cite{Hybridflow} exemplified this approach by integrating a discrete codebook-based stream with a continuous feature stream, effectively balancing perceptual quality and fidelity. HDCompression \cite{lu2025hdcompression} further advanced this idea by introducing diffusion models into the dual-stream architecture, enhancing the performance in image compression.

Building upon these advancements, we extend the dual-stream framework to the video domain and address the unique challenges introduced by the temporal dimension in our tokenization-based video compression approach. This adaptation not only improves spatiotemporal redundancy reduction, but also ensures robust, balanced performance in ultra-low bit rate scenarios.

\subsection{Tokenization for Video Compression}
\label{sec:video_compression}

In contrast to static images, the application of tokenization to video data is relatively underexplored. Although methods like MagViT \cite{Magvit} and TimeSformer \cite{TimeSformer} have extended token-based architectures to video generation and high-level recognition tasks, no prior work has investigated token-based video compression. 

This may be due to the inherent challenges mentioned above in applying tokenization to video data in Section \ref{sec:relatedTVR}. Compared to static images, the complex spatiotemporal visual content in videos demands either a significantly larger number of spatiotemporal tokens or a highly dimensional latent space for accurate modeling. This increased complexity substantially increases the number of bits required to effectively represent video-based TVR.

More recently, diffusion-based video tokenizers, such as Divot \cite{ge2025divot}, have been introduced to learn rich video representations for comprehension and generation. These approaches highlight the potential of tokenization for video understanding and high-fidelity synthesis. However, they primarily target medium-to-high bit rate regimes, where the emphasis is on reconstruction quality and generative flexibility. In contrast, our work focuses on ultra-low bit rate compression, positioning these advances as complementary to our study.

In particular, we address the unique challenges of the ultra-low bit rate setting, where efficient token usage and compact representations are very critical. To this end, we adopt the state-of-the-art Cosmos tokenizer \cite{CosmoSR}, whose temporally causal design makes it especially suitable for video compression. Building on Cosmos, we further introduce a masked token prediction mechanism that exploits temporal redundancy, enabling our framework to achieve highly efficient compression while preserving perceptual quality.

% To effectively address this challenge, we integrate the state-of-the-art (SOTA) Cosmos tokenizer \cite{CosmoSR}, which features an efficient temporally causal design. Furthermore, we introduce a masked token and prediction mechanism that aligns with Cosmos's temporally causal structure, effectively exploiting visual redundancy in videos for improved compression efficiency.  

\section{Methodology: Tokenized Video Compression}
\label{sec:TVR-main}

An overview of the full TVC pipeline is shown in Fig.~\ref{fig:network_structure}. We formulate video compression as a token-level selection and prediction problem and tackle the challenges of ultra-low bit rate scenarios through four key components

\begin{itemize}
    \item[1)] A dual-stream tokenizer that leverages Cosmos's spatiotemporal tokenizer to extract discrete and continuous token representations;
    \item[2)] Dedicated, entropy-efficient token compression pipelines for each stream, utilizing the checkerboard context models;
    \item[3)] A masked token prediction mechanism based on the spatiotemporal context, implemented through a lightweight Transformer;
    \item[4)] A multi-scale hierarchical pixel decoder that fuses the discrete and continuous streams to harness their complementary strengths.
\end{itemize}

This architecture enables the reconstruction of high-quality videos from highly sparse token representations, delivering efficient compression while preserving visual integrity and temporal consistency.

\subsection{Discrete and Continuous Tokenizers}
\label{sec:tokenizer}

As shown in Fig.~\ref{fig:network_structure}, our TVC framework tokenizes each video at the granularity of a group of pictures (GoP), allowing the model to capture both intra-frame and inter-frame dependencies. 
To represent videos compactly yet expressively, we extract two types of token streams from each GoP: a discrete token stream that encodes high-level semantics, and a continuous stream that preserves fine-grained spatial fidelity. Both are derived from the Cosmos tokenizer, which features a pretrained encoder-decoder architecture capable of producing both discrete and continuous token outputs. This dual-stream representation forms the foundation of our efficient and semantically rich video compression pipeline.

Following the notations in Cosmos \cite{CosmoSR}, let $x_i\!\in\!\newline\mathbb{R}^{(1+T)\times H \times W \times 3}$ denote the $i$-th GoP, where $H$, $W$, $1+T$ are height, width, and frame number, respectively. The tokenizer encoder utilizes wavelet transforms and spatiotemporal factorized 3D convolution to transform $x_i$ into a compact latent space. The spatiotemporal factorized causal self-attention mechanism effectively captures long-range spatiotemporal dependencies, resulting in a continuous latent token $y^C_i \!\in\! \mathbb{R}^{(1\!+\!\frac{T}{8}) \times \frac{H}{16} \times \frac{W}{16} \times d_C}$ retaining rich contextual information of $x_i$, where $d_C=16$ denotes the channel dimension of the continuous latent. To achieve a more compact representation in the discrete space, the discrete tokenizer incorporates the FSQ to further quantize the continuous feature into a discrete token map $y^D_i \!\in\! \mathbb{R}^{(1\!+\!\frac{T}{8}) \times \frac{H}{16} \times \frac{W}{16} \times d_D}$, where $d_D=6$ represents the channel dimension of the discrete tokens. The pixel decoder leverages both discrete and continuous tokens to reconstruct a high-quality GoP $\hat{x}_i\!\in\!\mathbb{R}^{(1+T)\times H \times W \times 3}$ that closely approximates the original $x_i$.

\subsection{Token Compression and Decompression}
\label{sec:tokencompression}

To enable ultra-low bit rate transmission, both the continuous latents $y^C_i$ and discrete code maps $y^D_i$ are compressed via CCMs and entropy coding.

\subsubsection{Continuous Stream}
We adopt a quantization-plus-entropy model pipeline based on the continuous CCM \cite{he2021checkerboard}. First, the $i$-th GoP $x_i$ is spatially downsampled to further reduce the size of the resulting continuous latent token $y^C_i$. It is quantized to $\hat{y}^C_i$ via uniform quantization, and $\hat{y}^C_i$ is compressed by an arithmetic encoder. The CCM predicts the Gaussian distribution parameters $\mu, \sigma$ for each token using both decoded context and a learned hyperprior $\hat{z}_i$, enabling efficient entropy modeling.

\subsubsection{Discrete Stream}

For the discrete token stream $y^D_i$, we compress the code map generated by FSQ rather than raw codebook indices, which offers a narrower dynamic range and thus better entropy coding precision. Discrete stream compression employs a CCM with a 3D spatiotemporal masking strategy to enable partial token transmission, i.e., only visible tokens are transferred.

During training we sample random masks following Ref.~\cite{MAGE}. During inference we use a fixed mask to keep the outputs stable. The fixed mask is built by revealing the first \(n_{\text{visible}}\) tokens in each mask interval of the flattened token sequence. The overall mask rate is

\begin{equation}
\text{Mask Rate} = 1 - \left({n_{\text{visible}}}/\text{Mask Interval}\right),
\end{equation}
\noindent where \(n_{\text{visible}}\) is the number of tokens that remain unmasked within one interval.
\noindent This 1D mask is reshaped into a 3D binary map $m\!\in\!\{0,1\}^{(1+\frac{T}{8})\!\times\!\frac{H}{16}\!\times\!\frac{W}{16}}$. The masked tokens are dropped and the visible ones are encoded via a CCM-based arithmetic encoder. As $y^D_i$ are integers from FSQ, this compression process of the visible tokens is lossless. The discarded tokens are later predicted in the decoder.

\subsubsection{Masking Consistency and Prediction Alignment}
To maintain semantic and structural alignment between the encoder and decoder, we adopt two consistency strategies. First, instead of using fixed constants as mask labels \cite{MAGE,Magvit,Hybridflow}, we dynamically initialize masked positions using values from the continuous stream $\hat{y}^C_i$. This improves both entropy modeling and Transformer-based prediction. Second, we apply the same 3D mask to the encoder-side input of the CCM, preserving spatial layout even for discarded tokens. This ensures alignment across the compression and reconstruction pipelines. Together, these techniques improve prediction fidelity and reduce mismatch in downstream token recovery.

\subsection{Masked Token Prediction}
\label{sec:maskpredictor}

The prediction of masked discrete tokens is performed by a decoder-only Transformer, as shown in Fig.~\ref{fig:network_structure}. To facilitate consistent spatiotemporal representation across token streams, we deliberately reapply the Cosmos tokenizer encoder to the discrete tokens prior to the Transformer prediction. This shallow encoder pass acts as a unified projection layer, aligning the structural characteristics of discrete tokens with those of the continuous stream. While seemingly redundant, this step ensures architectural symmetry, stabilizes early-stage Transformer training, and retains the inductive bias embedded in the original tokenization process.

 For compatibility with standard token prediction tasks, we convert the FSQ-generated code map back to its corresponding index map using the implicit codebook. Given a partially observed index map $y^{D(\text{index}),\text{mask}}_i$, where masked tokens have been dropped during compression, the Transformer reconstructs the complete token sequence $\hat{y}^{D(\text{index})}_i$ by attending to both intra-stream and cross-stream contexts, which is computed using Eq. (2):
 
{\small
\begin{eqnarray}
&& \hat{y}^{D(\text{index})}_i = \nonumber \\
&&\text{Transformer}(Q = y^{D(\text{index}),\text{mask}}_i, \ K = \hat{y}^{C}_i ,\ V = \hat{y}^{C}_i),
\end{eqnarray}
}

\noindent where $Q$, $K$, and $V$ represent the query, key, and value in the cross-attention mechanism, respectively. Here $y^{D(\text{index}),\text{mask}}_i$ contains the visible tokens of the discrete index stream, and $\hat{y}^{C}_i$ is the continuous latent stream that guides the prediction. The output $\hat{y}^{D(\text{index})}_i$ is the reconstructed full index sequence.

The Transformer comprises stacked spatiotemporal attention blocks, with learnable spatial and temporal positional embeddings to preserve the structure of token layouts. These layers capture both intra-frame dependencies and inter-frame correlations, enabling context-aware token reconstruction. Additionally, the cross-attention module leverages the continuous latent $\hat{y}^C_i$ as key and value (K,V) inputs, which provide guidance to improve prediction accuracy.

\subsection{Hierarchical Decoder Fusion}
\label{sec:decoder}

Inspired by the residual conditioning mechanism of ControlNet \cite{zhang2023controlnet}, our pixel decoder exploits the continuous token stream as a conditional control branch to guide the Cosmos tokenizer's pixel decoder for video reconstruction. In particular, latent features extracted from the continuous stream are injected into the main decoding path in a residual form at multiple spatial resolutions and are progressively fused with the discrete stream pixel decoder. This multi-scale residual fusion mechanism enables effective cross-stream information interaction, enhancing the pixel decoder’s contextual awareness. As a result, the model can better leverage the structurally preserving guidance from the continuous stream, leading to improved reconstruction accuracy and structural consistency.

\subsection{Training Pipeline}
\label{sec:training}

We adopt a four-stage training procedure to progressively optimize different components of the proposed TVC framework.

\noindent\textbf{Pretraining.}
We use the pretrained Cosmos discrete and continuous tokenizers~\cite{CosmoSR} to extract both $y_i^D$ and $y_i^C$. These tokenizers are kept frozen throughout the entire training process.

\noindent\textbf{Context model training.}
The CCMs for both streams are trained using an entropy loss as Eq. (3):
\begin{equation}
\mathcal{L}_{\text{bpp}} = \mathbb{E}_{x \sim p_x}\left[ -\log_2 p_{\hat{y}|\hat{z}}(\hat{y}_i|\hat{z}_i) - \log_2 p_{\hat{z}}(\hat{z}_i) \right],
\end{equation}
where $\hat{y}_i$ and $\hat{z}_i$ denote the quantized latent and hyper latent representations, respectively. 
Here $p_{\hat{y}|\hat{z}}$ and $p_{\hat{z}}$ are the conditional and prior probability models predicted by the context model. 
The term $\mathbb{E}_{x \sim p_x}$ is the expectation over the data samples $x$ drawn from the data distribution $p_x$. 
This objective corresponds to the expected bit cost under the predicted Gaussian distribution.
No distortion loss is applied at this stage, as the tokenizers are frozen.

\noindent\textbf{Transformer training.}
The Transformer decoder for reconstructing masked discrete tokens is trained based on a negative log-likelihood loss that is computed only over masked positions, which is computed using Eq. (4):
\begin{equation}
\mathcal{L}_{\text{recon}} = -\frac{1}{|\mathcal{M}|} \sum_{i \in \mathcal{M}} \log p_\theta(y_i^{D} \mid y^{D,\text{mask}}_i, \hat{y}^C_i),
\end{equation}
where $\mathcal{M}$ contains masked token indices and $p_\theta$ denotes the predicted probability distribution parameterized by the Transformer. 
Here $y_i^{D}$ is the ground truth discrete token at position $i$ and $y^{D,\text{mask}}_i$ is its masked version that keeps only the visible tokens. 
The term $\hat{y}^C_i$ is the continuous latent that provides cross stream guidance. 
Dropped tokens are recovered using both intra stream context and cross stream guidance.

\noindent\textbf{Pixel decoder fine-tuning.}
Finally, we fine-tune the ControlNet-style pixel decoder for video reconstruction. Latent features from the C-TVR stream are injected into the D-TVR decoder in a residual manner at multiple scales. The training loss combines pixel-wise and perceptual and it is computed using Eq. (5):
\begin{equation}
\mathcal{L}_{\text{pixel}} = \lambda_1 \cdot \mathcal{L}_{\text{L1}}(x, \hat{x}) + \lambda_2 \cdot \mathcal{L}_{\text{perc}}(x, \hat{x}),
\end{equation}
where $x$ and $\hat{x}$ denote the original and reconstructed GoPs, respectively, 
$\lambda_1$ and $\lambda_2$ are weighting coefficients, 
$\mathcal{L}_{\text{L1}}$ denotes the L1 loss, 
and $\mathcal{L}_{\text{perc}}$ represents the LPIPS loss, 
which together enhance visual fidelity by aligning reconstructed frames with both low-level structure and high-level semantics,

% TODO (will be fixed with rebuttal comments)
\subsection{Complexity Analysis}
\label{sec:complexity_analysis}

The proposed TVC is a holistic framework built upon TVR, which is inherently computationally efficient. It avoids redundant computation across modules and employs a single-pass inference process to handle each GoP for both the encoder and decoder. In contrast, traditional video codecs based on motion estimation and residual coding require the entire decoder-side process to be replicated within the encoder, which increases complexity. Existing neural video codecs adopt a piecemeal replacement strategy, substituting individual components with separate networks -- resulting in significant redundancy and high computational overhead. 

The modules in TVC are designed with computational efficiency in mind; for example, the Cosmos tokenizers and related components run several times faster than prior implementations. The CCM method is specifically optimized for fast learned image compression, offering SOTA computational efficiency. In the D-TVR stream, our masking mechanism significantly reduces the number of discrete tokens to be processed, while the C-TVR stream benefits from input downsampling, further minimizing its computational load. Additionally, the Transformer-based token predictor is lightweight by design. As a result, the integration of these efficient components within a single-pass inference process enables TVC to run efficiently.

As shown in Table \ref{tab:complexity_comparison}, we evaluate the computational complexity using an AMD EPYC 7513 CPU and a single NVIDIA L40s GPU with 1080p video inputs. Our method demonstrates efficiency, requiring 765 GMACs while achieving an encoding time of 214\,ms and a decoding time of 168\,ms, showing clear improvements over existing methods. For instance, DCVC-HEM\cite{DCVC-HEM} utilizes 3279 GMACs and requires 331\,ms for encoding and 236\,ms for decoding. Similarly, DCVC-DC\cite{DCVC-DC} operates at 2642 GMACs with 340\,ms for encoding and 259\,ms for decoding. Consequently, our model reduces multiply-accumulate (MAC) operations by approximately 77\% compared to DCVC-HEM\cite{DCVC-HEM} and 71\% compared to DCVC-DC\cite{DCVC-DC}, while also accelerating both encoding and decoding processes.

A detailed breakdown of our model’s complexity highlights its lightweight design. The Cosmos tokenizer accounts for 313 GMACs (and is frozen during training), while the trainable components — the checkerboard module (80 GMACs), the transformer predictor (207 GMACs), and CtrlNet (163 GMACs) — are all individually lean. This efficient design demonstrates the strength of our modular architecture, which not only preserves high reconstruction performance but also ensures practical deployment feasibility for high-resolution video compression.

We would also like to point out that one of the strengths of our framework is its flexibility to accommodate different tokenizers and decoder variants. Naturally, the MACs and runtime may vary depending on the specific modules used. For example, future advances in tokenization technology—an area developing rapidly—will directly translate to improved computational efficiency in our framework.

% \begin{table}[t]
%   \centering
%   \renewcommand{\arraystretch}{1.1}
%   \setlength{\tabcolsep}{1.5mm} 
%   \begin{tabular}{l c c c} \hline 
%                             & MAC (G)      & Encoding time (ms) & Decoding time (ms) \\ \hline
%     DCVC-HEM\cite{DCVC-HEM} & 3279     & 331       & 236       \\
%     DCVC-DC\cite{DCVC-DC}   & 2642     & 340       & 259       \\
%     Ours                    & 765      & 214       & 168       \\ \hline
%   \end{tabular}
%   \caption{Complexity comparison. Tested on single NVIDIA L40s GPU with using 1080p as input.}
%   \label{complexity comparison}
% \end{table}

\begin{table}[t]
  \centering
  \renewcommand{\arraystretch}{1.1}
  \setlength{\tabcolsep}{1.5mm} 
  \caption{Complexity comparison. Tested on single NVIDIA L40s GPU using 1080p input.}
  \label{tab:complexity_comparison}
  \vspace{0.3em}
  \begin{tabular}{l c c c} \hline
        & MAC & Encoding time & Decoding time \\ 
        & (G) & (ms)          & (ms)          \\ \hline
    DCVC-HEM\cite{DCVC-HEM} & 3279 & 331 & 236 \\
    DCVC-DC\cite{DCVC-DC}   & 2642 & 340 & 259 \\
    Ours                    & 765  & 214 & 168 \\ \hline
  \end{tabular}
  \vspace{0.5em}
\end{table}

% Through evaluation, our TVC runs about 2$\times$ faster than SOTA neural codecs like DCVC-DC \cite{DCVC-DC}, and runs more than 150$\times$ faster than conventional codecs like VTM \cite{VTM}. On average, the complete Encoder–Decoder pipeline of TVC runs at 0.3 sec/frame, compared to 50 sec/frame for VTM and 0.6 sec/frame for DCVC-DC, when evaluated on a single NVIDIA L40s GPU -- demonstrating TVC’s practical efficiency. Notably, TVC's concise architecture ensures stable and consistent computational complexity across different videos. This contrasts with traditional codecs, which exhibit significant variability in computation due to complex module selection and tool-switching mechanisms affected by video content.

% To process 1080p videos, with the AMD EPYC 7513 CPU and a single NVIDIA L40s GPU, our method achieves 214\,ms encoding and 168\,ms decoding time with only 765 GMACs - a 71\% computation reduction compared to DCVC-DC, which takes 340\,ms encoding / 259\,ms decoding time with 2642 GMACs. A detailed breakdown of our GMACs shows that the Cosmos Tokenizer contributes 313 GMACs (frozen during training), while the trainable components—checkerboard (80 GMACs), transformer predictor (207 GMACs), and CtrlNet (163 GMACs)—are all individually lightweight. This efficient design highlights the strength of our modular architecture, which not only preserves high reconstruction performance but also ensures practical deployment feasibility for high-resolution video compression.

\section{Experiments}
\label{sec:experiments}

% \begin{figure*}[t]
%   \includegraphics[width=1.0\linewidth]{rd_curve.png}
%   \caption{Rate-distortion (R-D) comparison (in terms of LPIPS; lower is better) across the UVG, MCL-JCV, and HEVC Class B datasets. Our TVC clearly outperforms conventional codecs such as VTM and HM in most cases. While existing neural codecs like DCVC-DC operate in a higher bit rate range, TVC is the first neural codec to achieve effective video reconstruction at ultra-low bit rates in the RGB colorspace, setting a new benchmark in this challenging scenario.}
%   \label{fig:rd_curve}
% \end{figure*}

\begin{figure*}[t]
  \includegraphics[width=1.0\linewidth]{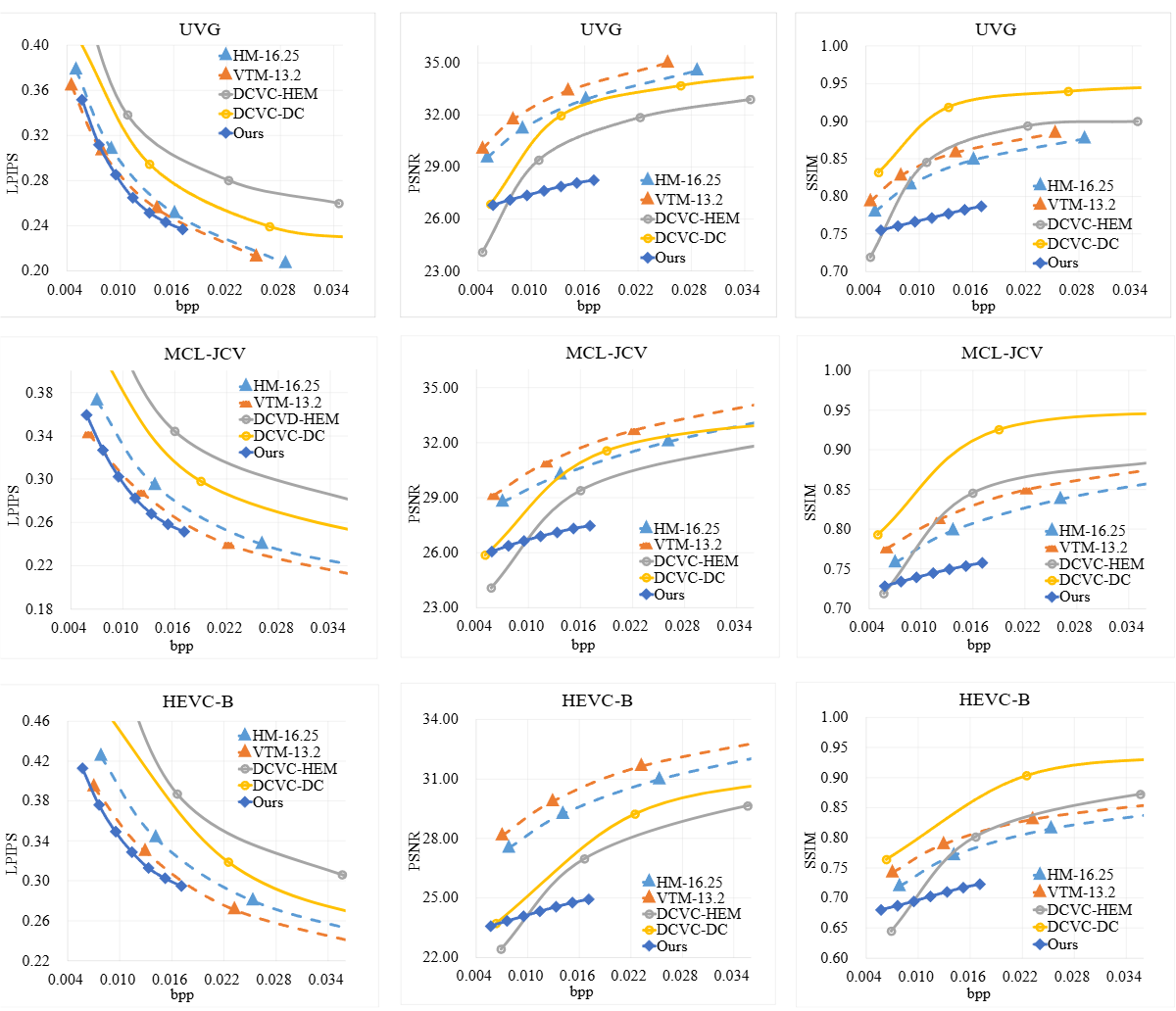}
  \caption{Rate–distortion (R–D) comparison on the UVG\cite{UVG}, MCL-JCV\cite{MCL-JCV}, and HEVC Class B\cite{HEVC} datasets in terms of LPIPS (lower is better), PSNR (higher is better), and SSIM (higher is better). TVC achieves lower LPIPS than both conventional and neural codecs in LPIPS at ultra-low bit rates. PSNR and SSIM results are also reported for completeness. }
  \label{fig:rd_curve}
  \vspace{-0.5em}
\end{figure*}

\begin{figure*}[t]
  \includegraphics[width=1.0\linewidth]{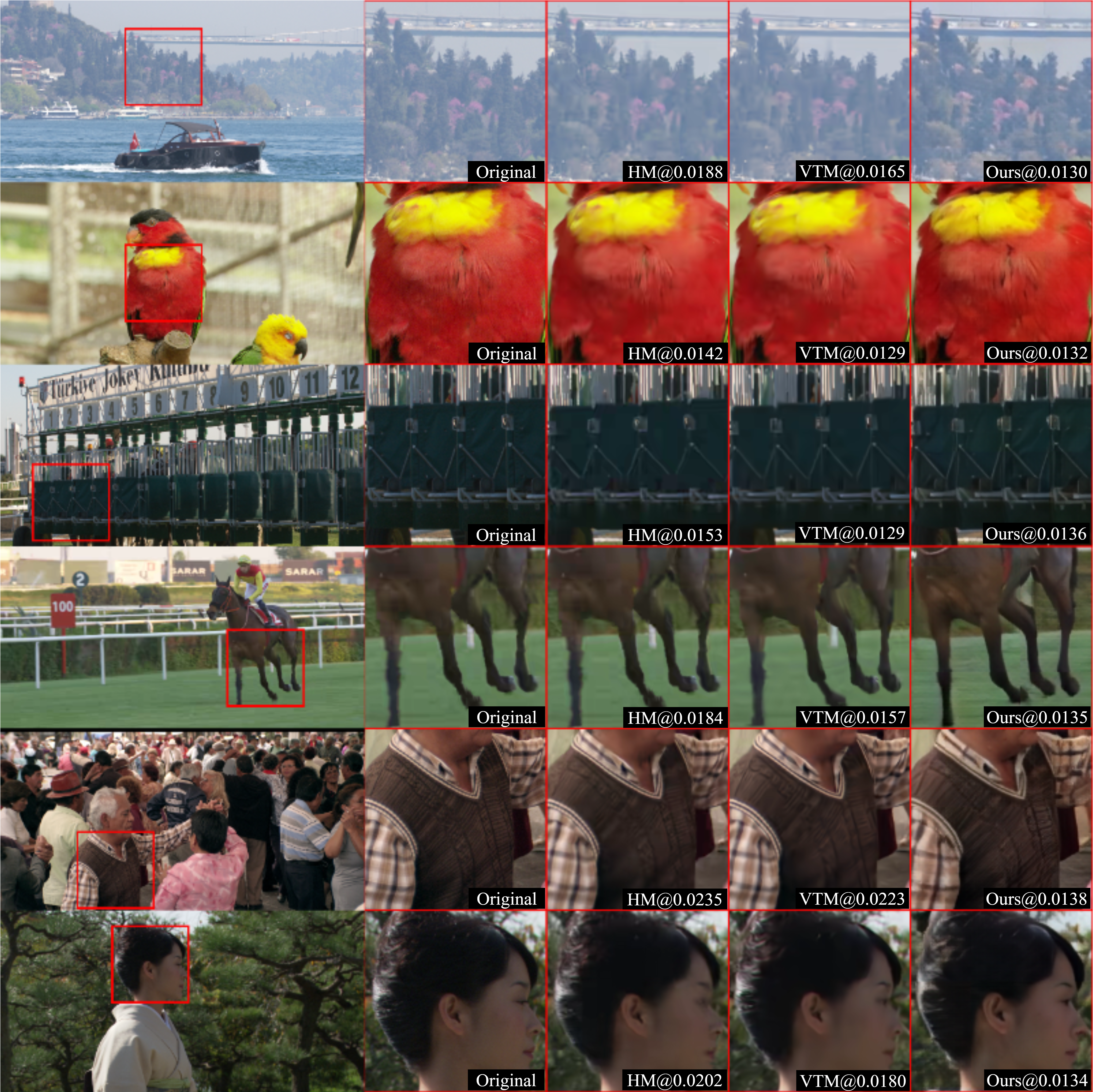}
  \caption{Qualitative comparison of reconstruction results across multiple test videos. The examples are taken from UVG\cite{UVG}, HEVC-B\cite{HEVC}, and MCL-JCV\cite{MCL-JCV}. TVC shows improvements over conventional codecs such as HM\cite{HM} and VTM\cite{VTM} in preserving both semantic structures and visual sharpness, even at lower bit rates. The actual bpp values are shown for each method. Our method provides reconstructions with fewer visual artifacts, such as blurring, ghosting, and blocking.}
  \label{fig:qualitative}
  \vspace{-0.5em}
\end{figure*}

\subsection{Experimental Settings}
\label{sec:settings}

\noindent\textbf{Datasets.} We trained our model on the Kinetics-600 training dataset \cite{K600-01, K600-02}. Specifically, video clips containing more than 300 frames and a resolution greater than $256\!\times\!256$ were selected from the training split. A centered $256\!\times\!256$ crop was extracted from each selected video to construct the training set. For evaluation, we followed prior works and conducted experiments on standard benchmarks, including HEVC Class B \cite{HEVC}, UVG \cite{UVG}, and MCL-JCV \cite{MCL-JCV}.\vspace{.3em}

\noindent\textbf{Implementation details.} We adopt the four-stage training strategy described in Section~\ref{sec:training}. In Stage~1, the model is initialized with publicly available pretrained weights. Stages~2–4 are trained sequentially with AdamW as the optimizer ($\beta_1 = 0.9$, $\beta_2 = 0.95$, and weight decay = 0.05). A half-cycle cosine learning rate schedule with linear warm-up is applied following common practice. Across these stages, training is performed for 10 epochs with 1 warm-up epoch, gradient accumulation = 1, a base learning rate of $5 \times 10^{-4}$ decayed to 0, and weight decay = 0.05. The only difference lies in batch size: Stage~2 uses 64, while Stages~3 and~4 use 16. The loss function in Eq.~(5) is weighted with $\lambda_1 = 1$ and $\lambda_2 = 4$. These values were chosen empirically to balance rate and distortion. Training was conducted on a single machine with 8$\times$ NVIDIA L40S GPUs, with a total training time of approximately 10 days. During inference, the fixed mask described in Section~\ref{sec:tokencompression} is employed. We set the mask interval to 7 and vary the number of visible tokens $n_{\text{visible}}$ from 1 to 7, which directly controls the rate–distortion trade-off and enables smooth adjustment of the RD curve. Bit rate is consistently measured in the RGB color space for both training and evaluation.

% \noindent\textbf{Compared baselines.} We compared our method against both traditional and learned video compression approaches. For traditional codecs, we included HM \cite{HM} and VTM \cite{VTM}, configured with the Low Delay setting. Among learned methods, we evaluated against DCVC-DC \cite{DCVC-DC}, a representative SOTA neural video codec. However, due to our focus on ultra-low bit rate settings, direct comparisons are limited, as existing methods like DCVC-DC do not operate effectively at such extreme bit rate levels.\vspace{.3em}

\noindent\textbf{Compared baselines.} We compared our method against both traditional and learned video compression approaches. For traditional codecs, we included HM \cite{HM} and VTM \cite{VTM}, configured with the low delay setting. Among learned methods, we follow Ref.~\cite{opendcvcs} to fine-tune DCVC-DC \cite{DCVC-DC} and DCVC-HEM \cite{DCVC-HEM} to operate in the ultra-low bit rate regime, enabling a fair comparison with overlapping bit rate ranges. We also attempted to fine-tune DVC \cite{DVC} and FVC \cite{FVC}; however, their achievable bit rates remained at or above 0.04 bpp, indicating that these methods cannot effectively operate under ultra-low bit rate conditions. Therefore, they were not included as baselines in our main comparison.\vspace{.3em}

\noindent\textbf{Metrics.} In the ultra-low bit rate regime, conventional metrics such as PSNR and MS-SSIM tend to prioritize structural fidelity while failing to capture perceptual realism. To better reflect human visual perception under extreme compression, we adopt LPIPS as our primary evaluation metric, due to its stronger alignment with perceptual quality and subjective visual preference.

% \subsection{Quantitative Performance}
% \label{sec:quantitative}

% Figure~\ref{fig:rd_curve} presents the quantitative rate-distortion (R-D) performance of various methods in terms of \textbf{LPIPS} on the UVG, MCL-JCV, and HEVC Class B datasets under ultra-low bit rate conditions. Compared to both traditional codecs and prior neural video compression approaches, our TVC consistently achieves superior perceptual quality, especially in the low-bit rate regime.

% Notably, TVC outperforms conventional codecs such as HM and VTM, and surpasses neural baselines like DCVC-DC across all datasets in most bit rate ranges, particularly in the sub-0.02 bpp regime. This performance gain stems from its structured dual-stream architecture and token prediction pipeline. The integration of the checkerboard entropy model and cross-attention transformer decoding enables more accurate and compact reconstructions, particularly in texture-rich or fast-motion scenes where other methods tend to degrade.

\subsection{Quantitative Performance}
\label{sec:quantitative}

Figure~\ref{fig:rd_curve} presents the quantitative rate–distortion (R–D) performance of various methods on the UVG\cite{UVG}, MCL-JCV\cite{MCL-JCV}, and HEVC Class B\cite{HEVC} datasets under ultra-low bit rate conditions. We report results in terms of LPIPS, PSNR, and SSIM. For fair comparison, both DCVC-DC\cite{DCVC-DC} and DCVC-HEM\cite{DCVC-HEM} were carefully fine-tuned to operate in the ultra-low bit rate regime.

TVC achieves lower LPIPS values than both traditional (HM\cite{HM} and VTM\cite{VTM}) and neural baselines (DCVC-DC\cite{DCVC-DC} and DCVC-HEM\cite{DCVC-HEM}) across our tested datasets. These results demonstrate that TVC delivers superior perceptual quality, particularly in the sub-0.02 bpp range where visual degradation is most pronounced.

Although perceptual metrics are the primary focus at ultra-low bit rates, TVC also maintains competitive or superior fidelity. In particular, at extremely low rates ($<$0.005 bpp), TVC even outperforms DCVC-HEM\cite{DCVC-HEM} in terms of both PSNR and SSIM. These results suggest that the dual-stream architecture can enhance perceptual quality while also preserving structural fidelity under extreme compression.

To better understand this behavior, we analyze how bits are allocated between the two streams. In our current design, the bit rate of the continuous stream (C-TVR) is fixed after the CCM stage, contributing approximately 0.003 bpp. This fixed cost provides a reliable lower bound on reconstruction fidelity. The overall R-D trade-off is then controlled by adjusting the mask rate of the discrete stream (D-TVR). Consequently, D-TVR dominates at ultra-low bit rates, while C-TVR offers a stable auxiliary signal to support reconstruction.

\subsection{Qualitative Performance}
\label{sec:quanlitative}

Figure~\ref{fig:qualitative} presents visual comparisons between TVC and both traditional and neural codecs across diverse video scenes. Despite operating at significantly lower bit rates (e.g., 0.013x vs. 0.018x), TVC retains fine details such as edges, textures, and semantic boundaries that are often smoothed out or distorted in HM\cite{HM} and VTM\cite{VTM} outputs.

Notably, the ControlNet-inspired fusion of continuous and discrete streams enables fine-grained texture reconstruction, while the Transformer-based token prediction improves global coherence and temporal consistency. These qualitative results validate our token-centric design, which effectively preserves both structure and semantics even under extreme compression.

\subsection{Ablation Study}
\label{ablation}

\begin{figure}[t]
  \includegraphics[width=1.0\linewidth]{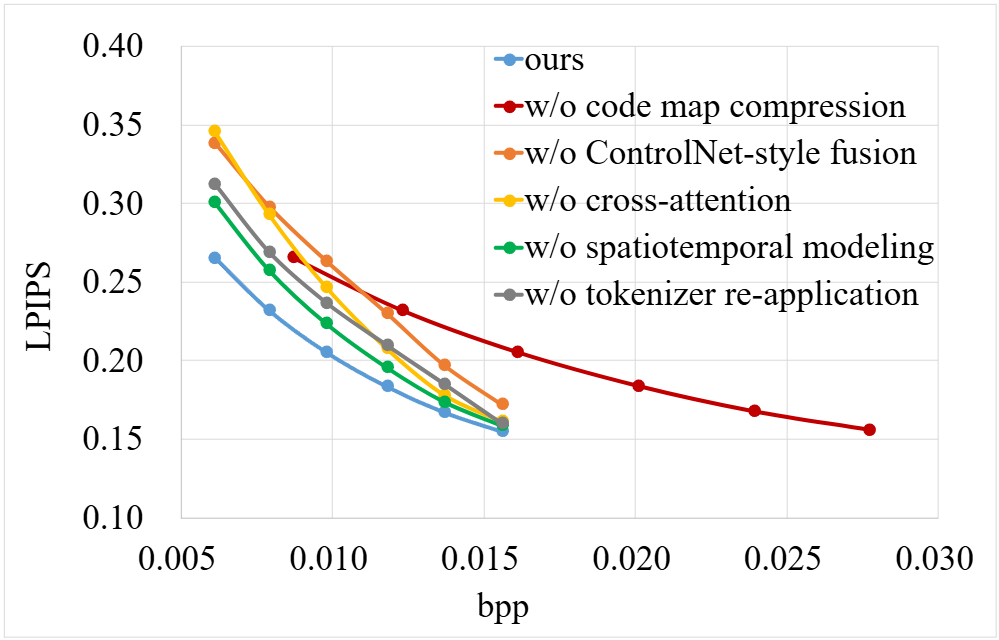}
    \caption{Ablation study evaluating the impact of key design components. Each variant disables a specific module from the full model, including code map compression, dual-stream fusion, cross-attention, and spatiotemporal modeling. Results show consistent performance drops across all ablations, highlighting the importance of each component.
}
  \label{fig:ablation}
\end{figure}

We conduct ablation experiments on the Kinetics-600 test set\cite{K600-01, K600-02} to assess the contribution of each architectural component. As shown in Fig.~\ref{fig:ablation}, removing any of the core modules leads to noticeable degradation in LPIPS, confirming their complementary roles.

\begin{itemize}
\item[1)] \textbf{Code map compression.} Disabling our code map compression leads to a consistent drop in rate-distortion performance—LPIPS increases by up to 0.05 at 0.01 bpp—due to inefficient entropy modeling over the high-cardinality index space.\vspace{.3em}
\item[2)] \textbf{Cross-attention.} Without the cross-attention mechanism, the perceptual quality degrades significantly, especially in semantically complex scenes. LPIPS worsens by 0.04 on average, indicating that continuous-token guidance is essential for accurate reconstruction of masked discrete tokens.\vspace{.3em}
\item[3)] \textbf{Spatiotemporal modeling.} Removing temporal attention and position encoding weakens the model’s ability to capture inter-frame dependencies, resulting in flickering artifacts and unstable predictions across frames. The LPIPS degradation becomes especially severe at lower bit rates.\vspace{.3em}
\item[4)] \textbf{ControlNet-style fusion.} The absence of dual-stream fusion causes the sharpest degradation in perceptual quality across all settings. LPIPS rises by over 0.05 at low bit rates, highlighting the critical role of stream interaction in preserving fine-grained structure.\vspace{.3em}
\item[5)] \textbf{Tokenizer re-application.} We also examine removing the re-application of the tokenizer encoder before Transformer prediction. In the variant w/o tokenizer re-application, discrete tokens are fed directly with mask tokens, leading to a modest performance drop of about 0.03 LPIPS. This step mainly serves to stabilize early training and helps the Transformer learn more robust predictions.
\end{itemize}

\subsection{Generalization to Alternative Tokenizers}
\label{sec:tokenizer_ablation}

To evaluate the generality of our framework, we replace the Cosmos tokenizer with OmniTokenizer~\cite{omnitoken} and re-train the downstream modules accordingly. While Cosmos represents the only open-source tokenizer with commercial-grade, state-of-the-art performance, OmniTokenizer~\cite{omnitoken} is the leading tokenizer in academic research. This experiment highlights the robustness of TVC under different tokenizer choices.

As shown in Table~\ref{tab:omni_lpips}, the dual-stream TVC design remains consistently effective when built on OmniTokenizer. At about 0.05 bpp, dual-stream TVC reduces LPIPS by 18.6\%/8.0\% compared with C-TVR/D-TVR on UVG\cite{UVG}, by 1.1\%/3.8\% on HEVC Class~B\cite{HEVC}, and by 6.8\%/3.2\% on MCL-JCV\cite{MCL-JCV}. These results confirm that TVC is not tied to a specific tokenizer, and is well-positioned to benefit from stronger tokenizers as tokenization technology continues to advance.

\begin{table}[t]
  \centering
  \renewcommand{\arraystretch}{1.1}
  \setlength{\tabcolsep}{3.5mm} 
  \caption{LPIPS comparison with OmniTokenizer\cite{omnitoken} at approximately 0.05 bpp. 
    Lower values indicate better perceptual quality. 
    Dual-stream TVC consistently outperforms both single-stream baselines. The best results are marked in bold.}
  \begin{tabular}{l c c c} \hline     
    Dataset & C-TVR & D-TVR & Dual-TVC \\ \hline
    UVG\cite{UVG}         & 0.3318 & 0.2935 & \textbf{0.2700} \\
    HEVC-B\cite{HEVC}     & 0.3605 & 0.3409 & \textbf{0.3281} \\
    MCL-JCV\cite{MCL-JCV} & 0.3507 & 0.3374 & \textbf{0.3267} \\ \hline
  \end{tabular}
  \label{tab:omni_lpips}
\end{table}

\subsection{Further Discussions}
\label{sec:futurediscussions}

Our TVC framework operates entirely in token space and is inherently modular, allowing for seamless integration with a wide range of tokenization strategies. While our current implementation builds upon Cosmos, recent work such as Ref.~\cite{sargent2025flow} introduces potential alternatives. The modular nature of TVC enables smooth integration with such emerging tokenizers. TVC can adaptively evolve in tandem with future advances in the visual tokenization techniques.

% Looking ahead, several promising directions remain to be explored. One is the development of content-adaptive masking, which would allow the model to allocate prediction efforts more effectively across spatial and temporal regions, potentially improving rate–distortion performance. Another is dynamic bitstream allocation between discrete and continuous streams, which could provide finer control over bit rate distribution and better adapt to varying video characteristics. Beyond these, we also envision extending the pipeline with class-conditional tokenization, multimodal alignment, and instruction-driven generation backbones. These advancements could enable personalized and task-specific compression solutions, tailored to applications such as video editing and synthesis.

Looking ahead, there are several promising directions that could further enhance the performance of our framework.

One is the development of content-adaptive masking, which would allow the model to allocate prediction efforts more effectively across spatial and temporal regions according to video content. To make this feasible, two main challenges need to be addressed: 1) designing reliable importance maps for spatiotemporal tokens across diverse video content, and 2) transmitting the mask pattern with minimal bit rate overhead. Solving these challenges has the potential to unlock significant performance gains for our framework.

Another direction is to extend the bit rate range with dynamic bit allocation between the discrete and continuous streams. Our current design is tailored for the ultra-low bit rate regime, where D-TVR dominates to maintain perceptual quality while C-TVR plays a supporting role by providing fidelity cues. In this setting, the bit rate of C-TVR is fixed after CCM training, and the overall rate–distortion trade-off is controlled by adjusting the mask rate of D-TVR during inference. As bit rate increases, however, this allocation strategy would need to adapt, with C-TVR contributing more bits to enhance fidelity and D-TVR providing complementary perceptual detail. Determining the optimal balance between perceptual quality and fidelity under different bit rate budgets and content types, and designing mechanisms to allocate bits accordingly are two important problems for future exploration.

Beyond these directions, we also envision extending the pipeline with class-conditional tokenization, multimodal alignment, and instruction-driven generation backbones. Such advancements could open the door to personalized and task-specific compression solutions, tailored to downstream applications such as video editing, content adaptation, and controllable video synthesis.

% Looking ahead, we envision extending the compression pipeline with class-conditional tokenization, multimodal alignment, and instruction-driven generation backbones. These advancements could enable personalized and task-specific compression solutions, tailored to applications like video editing and synthesis.

\section{Conclusion}
\label{conclusion}

We present TVC, a video compression framework that implements a fully tokenized pipeline designed for operation at ultra-low bit rates. By framing compression as a process of token selection and prediction, TVC offers an alternative to conventional approaches based on motion estimation and residual coding, providing a more semantic and flexible way to represent video content.

The dual-stream architecture separates semantic content from fidelity details, creating a representation space that can adapt across different bit rate regimes. The integration of causal attention mechanisms, checkerboard modeling, and ControlNet-inspired fusion illustrates how a semantic-first, perception-driven approach can achieve effective compression at rates as low as 0.01 bpp.

An additional contribution of TVC is to bring video compression closer to modern generative modeling techniques by shifting the perspective from pixel-based compression to token-based representation. This perspective may encourage future work that connects compression with downstream applications such as content retrieval, video synthesis, and interactive systems.

% % For one-column wide figures use
% \begin{figure}[t]
%   \includegraphics[width=1.0\linewidth]{vi_example.pdf}
%   \caption{One-column figure. Please write figure caption here.}
%   \label{fig:1}       % Give a unique label
% \end{figure}
% %
% % For two-column wide figures use
% \begin{figure*}[t]
%   \includegraphics[width=1.0\linewidth]{vi_example.pdf}
%   \caption{Two-column figure. Please write figure caption here.}
%   \label{fig:2}       % Give a unique label
% \end{figure*}
% %
% % For one-column wide table use
% \begin{table}[t]
%   \centering
% 	\renewcommand{\arraystretch}{1.1}
% 	\setlength{\tabcolsep}{8.5mm}
%   \begin{tabular}{lll} \hline 
%     first & second & third  \\ \hline
%     number & number & number \\
%     number & number & number \\ \hline
%   \end{tabular}
%   \caption{One-column table. Please write table caption here.}
%   \label{tab:1}       % Give a unique label
% \end{table}
% %
% % For two-column wide figures use
% \begin{table*}[t]
%   \centering
% 	\renewcommand{\arraystretch}{1.3}
% 	\setlength{\tabcolsep}{22mm}
%   \begin{tabular}{lll} \hline 
%     first & second & third  \\ \hline
%     number & number & number \\
%     number & number & number \\ \hline
%   \end{tabular}
%   \caption{Two-column table. Please write table caption here.}
%   \label{tab:2}       % Give a unique label
% \end{table*}

\begin{small}
\vspace{.15in}  
\noindent \textbf{Abbreviations}

\noindent
AE: autoencoder;
Bpp: bit per pixel;
CCM: checkerboard context model;
C-TVR: continuous tokenized visual representation;
D-TVR: discrete tokenized visual representation;
FSQ: finite scalar quantization;
GAN: generative adversarial network;
GoP: group of pictures;
LIC: learned image compression;
LPIPS: learned perceptual image patch similarity;
LVC: learned video compression;
MACs: multiply accumulate operations;
MAGE: masked generative encoder;
MEMC: motion estimation and motion compensation;
MS-SSIM: multi-scale structural similarity;
PSNR: peak signal-to-noise ratio;
RD: rate-distortion;
TVC: tokenized video compression;
TVR: tokenized visual representation;
VAE: variational autoencoder;
VQGAN: vector quantized generative adversarial network.

\vspace{.15in}  
\noindent \textbf{Data Availability}

\noindent The public datasets used and analyzed during this study are available from the following sources:
\begin{itemize}
    \item[1)] \textbf{Kinetics dataset}, \url{https://github.com/cvdfoundation/kinetics-dataset}.
    \item[2)] \textbf{HEVC test sequences}, \url{http://ftp.kw.bbc.co.uk/hevc/hm-10.0-anchors/bitstreams/}.
    \item[3)] \textbf{MCL-JCV dataset}, \url{https://huggingface.co/datasets/uscmcl/MCL-JCV_Dataset}.
    \item[4)] \textbf{UVG dataset}, \url{https://ultravideo.fi/dataset.html}.
\end{itemize}
%
%The datasets generated during and/or analyzed during the current study are 
%available from the corresponding author on reasonable request.

\vspace{.15in}  
\noindent \textbf{Competing Interests}

% Authors are required to disclose financial or non-financial interests 
% that are directly or indirectly related to the work submitted for publication.
\noindent Wei Wang and Wei Jiang are employees of Futurewei Technologies Inc. During the period this research was conducted, Lebin Zhou and Cihan Ruan were research interns at Futurewei Technologies Inc. Nam Ling is a professor at Santa Clara University and the Ph.D. advisor for Lebin Zhou and Cihan Ruan. The remaining author, Zhenghao Chen, declares no competing interests. This study received support from Futurewei Technologies Inc.

\vspace{.15in}  
\noindent \textbf{Author Contributions}
% For research articles authors may use CRediT taxonomy:
% Conceptualization: [full name], …; 
% Methodology: [full name], …; 
% Formal analysis and investigation: [full name], …; 
% Writing - original draft preparation: [full name, …]; 
% Writing - review and editing: [full name], …; 
% Funding acquisition: [full name], …; 
% Resources: [full name], …; 
% Supervision: [full name],….
% %
% For review articles where discrete statements are less applicable a 
% statement should be included who had the idea for the article, 
% who performed the literature search and data analysis, 
% and who drafted and/or critically revised the work.

\noindent The study's conception and methodology were designed by Lebin Zhou, Wei Jiang, and Wei Wang. Lebin Zhou performed the data collection, investigation, and validation, with substantial support and guidance from Wei Jiang and Wei Wang. The formal analysis, data curation, and visualization were conducted by Lebin Zhou, Cihan Ruan, Wei Jiang, and Wei Wang. The original draft of the manuscript was also prepared by this group. Nam Ling and Zhenghao Chen provided overall supervision, contributed to the analysis through constructive discussions, and critically reviewed and edited the manuscript. All authors have read and approved the final manuscript.

\vspace{.15in}  \noindent \textbf{Funding}

\noindent This research was supported by Futurewei Technologies Inc. through funding for a research collaboration and student internships.

% acknowledgments part
% Acknowledgments of people, grants, \etc. 
% you can acknowledge any support given which is not covered by the author 
% contribution or funding sections. 
% This may include administrative or technical support.

% Funding: Please add “No funding was received to assist with the preparation of this manuscript.” ; 
% “No funding was received for conducting this study.” 
% Or “This research was funded by [NAME OF FUNDER], grant number […]” ; 
% This work was supported by […] (Grant numbers […] and […]). 
% Check carefully that the details given are accurate and use the standard spelling 
% of funding agency names at https://search.crossref.org/funding.
\vspace{.15in}  \noindent \textbf{Acknowledgments}

\noindent The authors would like to thank the anonymous reviewers for their valuable comments. The authors also gratefully acknowledge the High-Performance Computing (HPC) resources provided by Futurewei Technologies Inc. for conducting the experiments in this study.
\end{small}

% BibTeX from reference.bib
% \bibliographystyle{unsrt}
\bibliographystyle{vi}
\bibliography{reference}

\end{document}